\documentclass[a4paper]{article}

\usepackage{itgspeech2012}    
\usepackage{times}            
\usepackage[english]{babel}   
\usepackage[utf8]{inputenc}		
\usepackage[T1]{fontenc}			
\usepackage[sort&compress,numbers]{natbib}	
\usepackage{amsmath,amssymb}
\usepackage{graphicx}
\usepackage{xcolor}
\usepackage{booktabs}
\usepackage{tabularx}
\usepackage{hyperref}

\usepackage{pgf}
\usepackage{pgffor}
\usepgflibrary{plothandlers}

\graphicspath{{images/}}

\title{Declipping of Speech Signals Using Frequency Selective Extrapolation}

\author{Markus Jonscher, Jürgen Seiler, André Kaup}

\address{Multimedia Communications and Signal Processing, University of Erlangen-Nuremberg, Cauerstr. 7, 91058 Erlangen\\
  Email: \texttt{\{jonscher,seiler,kaup\}@LNT.de}\\
  Web: \texttt{www.lms.lnt.de}}

\begin{document}
\maketitle

\begin{abstract}
\vspace*{-0.1cm}
The reconstruction of clipped speech signals is an important task in audio signal processing to achieve an enhanced audio quality for further processing. In this paper, Frequency Selective Extrapolation (FSE), which is commonly used for error concealment or the reconstruction of incomplete image data, is adapted to be able to restore audio signals which are distorted from clipping. For this, FSE generates a model of the signal as an iterative superposition of Fourier basis functions. Clipped samples can then be replaced by estimated samples from the model. The performance of the proposed algorithm is evaluated by using different speech test data sets. Compared to other state-of-the-art declipping algorithms, this leads to a maximum gain in SNR of up to $3.5$ dB and an average gain of $1$ dB.
\end{abstract}

\vspace*{-0.1cm}
\section{Introduction}
\label{sec:intro}

For applications like speech recognition or further processing it is important to have undistorted speech signals with the best possible audio quality. 
This is also true for vocal tracks in music signals, since a modern trend in music business aims at producing music at the highest possible loudness level in order to differ from other artists or competitors. Due to the fact that people respond more easily to louder audio stimuli, this leads to the illusion of {\it{louder equals better}}. This phenomenon was first investigated by Fletcher and Munson \cite{Fletcher1933} who came up with the equal-loudness contour. They figured out that people perceive a mono-frequent tone with a fixed sound pressure level differently loud over the whole frequency spectrum. 
For higher sound pressure levels, this difference is getting smaller which means that music played at high volume is perceived more linear than it actually is and is therefore perceived as \emph{better}. Since digital signals cannot output values higher than the maximum representable value, it is not possible to increase the loudness beyond this limit. Instead, the quiet parts of a signal are amplified which leads to a subjective perception of an higher loudness level but also to a reduction of the signal's dynamic range. As a consequence, clipping may occur if the amplification is done in an improper way.

\begin{figure}
	\centering
	\def\svgwidth{\columnwidth}
	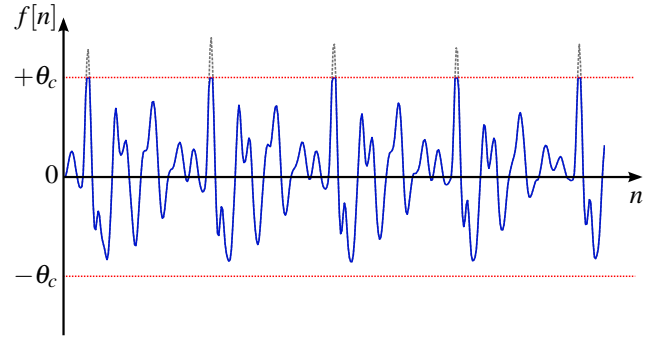
	\vspace*{-0.7cm}
	\caption{Clipped audio signal $f[n]$ at a certain clipping threshold $\theta_c$.}
	\label{fig:clipped_signal}
	\vspace*{-0.4cm}
\end{figure}

Clipping in general may occur when the range of all representable values is exceeded. This might be the case for example when a microphone for acquiring the data has some limitations on the range of values that can be measured. Clipping may also occur when converting a signal from the analog domain to the digital domain or during post-processing. When a signal is clipped, all its samples that lie beyond the maximum representable value, which is referred to as the clipping threshold $\theta_c$, are mapped onto $\pm\theta_c$, as it can be seen exemplarily in Figure~\ref{fig:clipped_signal}. In this paper, only digital hard clipping is regarded whereas in analog systems, soft clipping is very common and may not be as harmful as hard clipping. Since the information that is contained in the peaks which are digitally clipped is completely eliminated, such hard clipped signals cannot be recovered easily to their original values.
Declipping methods are therefore necessary to achieve an enhanced level of audio quality for listeners, audio applications and further processing of such signals.

In past and recent research, several approaches have been proposed which are able to restore clipped audio signals up to a certain degree. All of these approaches are just approximations of the original signal, however, the subjective quality is usually improved.
A promising method to reconstruct audio data which has been distorted due to clipping is the Audio Inpainting Framework proposed in \cite{Adler2012}. In this framework, the clipped audio samples are treated as missing samples and the signal is then decomposed into overlapping time-domain frames. The reconstruction is formulated as an inverse problem per audio frame and solved by using the Orthogonal Matching Pursuit algorithm in combination with a discrete cosine or a Gabor dictionary.
In \cite{Janssen1986}, an adaptive algorithm for the reconstruction of lost samples in discrete-time signals is presented by Janssen. It is based on linear prediction and for the region embedding the missing samples, a single auto regressive model is considered where the set of auto regressive coefficients and the set of missing samples are alternately estimated by an expectation maximization-like iterative algorithm. Declipping is not mentioned as an explicit application for this method, however, it is naturally suited for this problem.
Another approach \cite{Abel1991} performs an extrapolation based on the knowledge of unclipped samples and the value range that clipped samples can take.
A more recent declipping algorithm has been proposed in \cite{Defraene2013}. It is jointly based on compressed sensing and the declipping problem is formulated as a sparse signal recovery problem. Additionally, the knowledge of well-known properties of human auditory perception is used for reconstruction.
Moreover, there are many open source or commercially available declipping solutions: for example a plug-in for Audacity named \emph{ClipFix} which is based on cubic interpolation or a complete audio restoration toolbox with special declipping functions like \mbox{\emph{iZotope RX}}.

In this paper, Frequency Selective Extrapolation (FSE) \cite{Seiler2010}, which is normally used for error concealment or the reconstruction of non-regular sampled image data \cite{Schoeberl2011a}, is adapted to the problem of reconstructing clipped audio signals. FSE iteratively generates a model of the signal as a superposition of Fourier basis functions and clipped samples in the distorted signal are then replaced by estimated samples from the model. The performance of the proposed algorithm is evaluated by using different speech data sets and is then compared to the Orthogonal Matching Pursuit algorithm using Gabor dictionary (OMP-G) \cite{Adler2012} and Janssen's method \cite{Janssen1986}. 

The paper is organized as follows: The next section covers the basic principle of Frequency Selective Extrapolation and how it can be used for the declipping of distorted audio signals. Section~\ref{sec:results} presents the experimental results and Section~\ref{sec:conclusion} concludes this paper.

\setlength{\emergencystretch}{0.0em}
\section{Frequency Selective Extrapolation}
\label{sec:fse}

\begin{figure}
	\hspace*{-0.3cm}
	\input{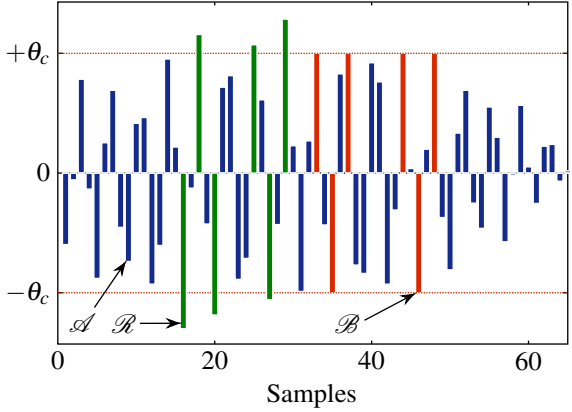}%
	\vspace*{-0.7cm}
	\caption{Extrapolation areas $\mathcal{L}$ includes the support area $\mathcal{A}$, the loss area $\mathcal{B}$, and the area $\mathcal{R}$ of already extrapolated samples.}
	\label{fig:extrapolation_areas}
	\vspace*{-0.4cm}
\end{figure}

Frequency Selective Extrapolation (FSE) \cite{Seiler2010} is usually utilized in error concealment or the reconstruction of incomplete image data and is now adapted to the one-dimensional problem of declipping audio signals. Therefore, the following scenario is regarded: a distorted audio signal $f[n]$ with the temporal coordinate $n$ is given and the clipped samples are to be reconstructed. All samples that lie beyond a certain clipping threshold $\theta_c$ are considered to be lost, that is to say that declipping is formulated as an extrapolation problem. These missing samples are gathered in the so called loss area $\mathcal{B}$. All samples with valid information are grouped into the support area $\mathcal{A}$ and those positions which contain all samples that have already been reconstructed are depicted by area $\mathcal{R}$. The areas $\mathcal{A}$, $\mathcal{B}$, and $\mathcal{R}$ are disjoint and the extrapolation area $\mathcal{L}$ is regarded as the union $\mathcal{A}\cup\mathcal{B}\cup\mathcal{R}$. All of these sub-areas are illustrated in Figure~\ref{fig:extrapolation_areas}.
\begin{figure}
	\vspace*{0.45cm}
	\input{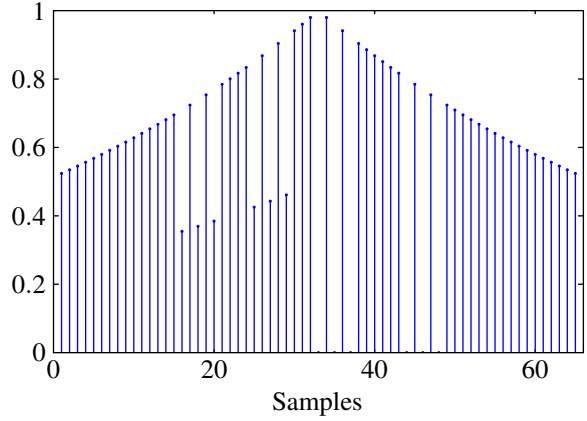}%
	\vspace*{-0.1cm}
	\caption{Weighting Function used for controlling the influence of neighboring samples.}
	\label{fig:weighting_function}
	\vspace*{-0.05cm}
\end{figure}
FSE, however, is a block-based algorithm and the missing parts of the signal are processed sample-wise. Therefore, the extrapolation area $\mathcal{L}$ consists of one sample that is currently processed surrounded by a certain amount of support samples on both sides.
Since the original signal $s[n]$ is only theoretically available, the measured signal
\begin{equation}
	f[n] = s[n]\cdot b[n]
\end{equation}
is regarded, where the original signal is masked by the masking function
\begin{equation}
		b[n] = 
		\begin{cases} 
			1 & \forall\ n\in\mathcal{A}, \mathcal{R} \\
			0 & \forall\ n\in\mathcal{B}
		\end{cases}.
	\label{eq:masking_function}
\end{equation}

\setlength{\emergencystretch}{0.5em}
FSE tries to estimate the signal $s[n]$ as exactly as possible by generating the sparse signal model $g[n]$ from the measured signal $f[n]$. Since all valid samples should be kept, only missing samples are replaced by corresponding samples from the model $g[n]$. The estimated signal
\begin{equation}
		\hat{s}[n] = 
		\begin{cases} 
			f[n] & \forall\ n\in\mathcal{A} \\
			g[n] & \forall\ n\in\mathcal{B}
		\end{cases}
	\label{eq:reconstructed_signal}
\end{equation}
can therefore be obtained by keeping all valid information from $f[n]$ and replacing all missing samples with samples from $g[n]$. FSE iteratively generates the sparse signal model
\begin{equation}
	g[n] = \sum_{k\in\mathcal{K}}\hat{c}_k\varphi_k[n]
\end{equation}
as a superposition of Fourier basis functions
\begin{equation}
	\varphi_k[n] = e^{j \frac{2\pi}{N}kn}
\end{equation}
weighted by the expansion coefficients $\hat{c}_k$. The indices of all considered basis functions are summarized in the set $\mathcal{K}$. In every iteration, one basis function is selected, its corresponding weight estimated and added to the model.

As proposed in \cite{Meisinger2004}, an isotropic weighting function $w[n]$ is used to control the influence that each sample has on the model generation. The weighting function
\begin{equation}
		w[n] = 
		\begin{cases} 
			\hat{\rho}^{(n-\frac{N-1}{2})} & \forall\ n \in \mathcal{A} \\
			\delta\hat{\rho}^{(n-\frac{N-1}{2})} & \forall\ n \in \mathcal{R} \\
			0 & \forall\ n \in \mathcal{B}
		\end{cases}
	\label{eq:weighting_function}
\end{equation}
is also divided into the three sub-areas mentioned above. Samples from the loss area $\mathcal{B}$ are neglected for the model generation. The decay of the weighting function in areas $\mathcal{A}$ and $\mathcal{R}$ is controlled by the decay factor $\hat{\rho}$. All already extrapolated samples in area $\mathcal{R}$ are additionally weighted by the weighting factor $\delta$. An example of a possible weighting function is illustrated in Figure~\ref{fig:weighting_function}.

The model to be generated is initialized to zero and will then be built iteratively by selecting one basis function in every iteration which is then added to the model generated so far. Therefore, in the generation process, the approximation residual
\begin{equation}
	r^{(\nu-1)}[n] = f[n] - g^{(\nu-1)}[n]
\end{equation}
from the previous iteration is projected onto all basis functions, where the weighting function is again used for controlling the influence of each sample on the model generation. This weighted projection leads to the projection coefficients
\begin{equation}
	p_k^{(\nu)} = \frac{\sum\limits_{n\in\mathcal{L}}{r^{(\nu-1)}[n]\,\varphi_{k}^{\ast}[n]\,w[n]}}{\sum\limits_{n\in\mathcal{L}}{\varphi_{k}^{\ast}[n]\,w[n]\,\varphi_{k}[n]}},\ \forall\, k.
\end{equation}
Afterwards, the basis function gets selected that minimizes the weighted distance between the residual and the weighted projection onto the corresponding basis function. The index of the basis function to be selected is determined by
\begin{equation}
	u^{(\nu)} = \underset{k}{\arg\!\max} \left( \left| p^{(\nu)}_k \right|^2 \sum\limits_{n\in\mathcal{L}}{\varphi_{k}^{\ast}[n]\,w[n]\,\varphi_{k}[n]} \right).
\end{equation}
To cope with the non-orthogonality of the basis functions, Orthogonality Deficiency Compensation (ODC) \cite{Seiler2007} is applied. Hence, the weight of the basis function results in
\begin{equation}
	\hat{c}_{u^{(\nu)}} = \gamma\,p_{u^{(\nu)}},
\end{equation}
where $\gamma$ is the so called ODC factor. After one basis function has been selected and its corresponding weight been estimated, the signal model is updated by
\begin{equation}
	g^{(\nu)}[n] = g^{(\nu-1)}[n] + \hat{c}_{u^{(\nu)}} \, \varphi_{u^{(\nu)}}[n].
\end{equation}
It is ensured that the reconstructed samples always lie between the clipping level and the maximum representable value.
Additionally, an optimized processing order similar to \cite{Seiler2011} is applied. Instead of processing the missing samples in a strictly linear order, those samples are processed first which have the most valid information in their neighborhood. This leads not only to an enhanced reconstruction quality but it gives also the possibility to process the model generation in parallel. For a faster computation, FSE can be carried out completely in the frequency domain.

\renewcommand{\arraystretch}{1.2}
\setlength{\tabcolsep}{6mm}
\begin{table}
	\caption{Set of parameters that are used by FSE for declipping audio signals.}
	\label{tab:fse_parameters}
	\centering
	\begin{tabularx}{\columnwidth}{lr}
		\toprule
		ODC Factor $\gamma$                          & $1.25$ \\
		Weighting Function Decay Factor $\hat{\rho}$ & $0.99$ \\
		Weighting Factor $\delta$                    &    $1$ \\
		Support Samples				                 & $1000$ \\
		FFT Size                                     & $2048$ \\
		Max. Iterations                              & $1500$ \\ \bottomrule
	\end{tabularx}
	\vspace*{-0.18cm}
\end{table}

\vspace*{0.2cm}
\section{Experimental Results}
\label{sec:results}
\vspace*{0.1cm}

For measuring the performance of the proposed algorithm and for comparative evaluations, different speech test data sets are used. The first test set \cite{Cooke2006} is named GRID and consists of $10$ both male and female speech signals which have been randomly chosen from the whole data set. Each test signal is around $1.5$ seconds long and sampled at $25$ kHz. This set is mainly used for training the different parameters of FSE. The second test set also consists of $10$ speech signals which are each 5 seconds long and sampled at $16$ kHz. These sound files are part of the freely available material of the $2008$ Signal Separation Evaluation Campaign (SISEC) \cite{Vincent2012}
and include both male and female speech from different speakers. This set is used for the comparison with other state-of-the-art declipping algorithms. Each original test signal is first normalized so that the maximum amplitude is $1$ and then artificially clipped with successive clipping levels from $0.5$ to $0.9$.

For comparing FSE with other state-of-the-art declipping algorithms, Orthogonal Matching Pursuit using Gabor dictionary (OMP-G) and Janssen's method have been chosen. The authors from \cite{Adler2012} provide their Audio Inpainting Toolbox. This MATLAB toolbox includes both the OMP-G algorithm and also Janssen's method.

To assess the performance of the reconstruction quality, the signal-to-noise ratio defined by
\begin{equation}
 \text{SNR}_{\mathrm{miss}} = 10\log_{10}\left( \frac{\sum\limits_{n\in\mathcal{B}}{s[n]^2}}{\sum\limits_{n\in\mathcal{B}}{(s[n] - \hat{s}[n])^2}} \right)
\label{eq:snr_miss}
\end{equation}
is used in all considered scenarios. $s[n]$ is the original undistorted audio signal and $\hat{s}[n]$ the reconstructed declipped signal. The SNR is only evaluated on sample positions where clipping occurred. 

To adapt FSE to be able to declip audio data, several parameters of FSE have to be evaluated. Therefore, the GRID test data set is used for training these parameters. For each clipping level, all artificially clipped audio signals are reconstructed by FSE, OMP-G, and Janssen's method respectively and an average of all resulting SNR values is calculated. In the end, for each clipping level and each declipping algorithm, there is one average SNR value. In Table~\ref{tab:fse_parameters}, one can see the parameters that give a good reconstruction quality and which are used for comparing FSE with OMP-G and Janssen's method on the SISEC test set.

\renewcommand{\arraystretch}{1.2}
\setlength{\tabcolsep}{5.5mm}
\begin{table}
	\caption{Runtime examples for FSE, OMP-G and Janssen's method.}
	\label{tab:time_measurements}
	\centering
	\begin{tabularx}{\columnwidth}{c|ccc}
		\toprule
		$\theta_c$ &  OMP-G   &  FSE   &  Janssen  \\ \midrule
		  $0.5$    & $495.74$ s & $13.40$ s & $0.76$ s \\
		  $0.6$    & $357.41$ s & $5.32$ s & $0.49$ s  \\
		  $0.7$    & $266.21$ s & $2.89$ s & $0.33$ s  \\
		  $0.8$    & $152.17$ s  & $1.65$ s & $0.20$ s  \\
		  $0.9$    & $134.98$ s  & $0.72$ s & $0.16$ s  \\ \bottomrule
	\end{tabularx}
	\vspace*{-0.3cm}
\end{table}

\begin{figure*}
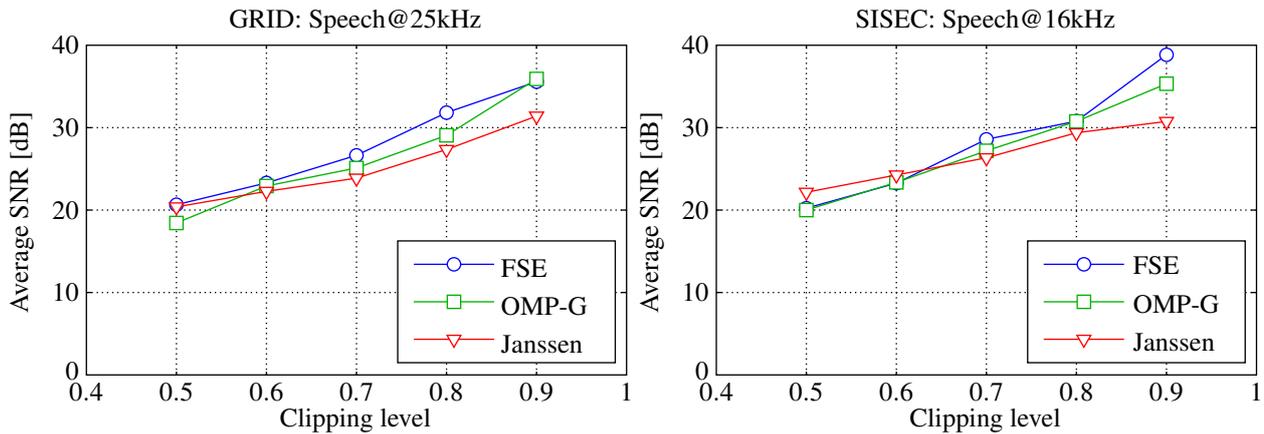

	\input{images/eva_grid.pgf}%
	\input{images/eva_sisec.pgf}%
	\vspace*{-0.22cm}
	\caption{Comparison with other existing declipping algorithms: Performance of FSE on the GRID data set (left plot) and on the SISEC data set (right plot).}
	\label{fig:results}
	\vspace*{-0.3cm}
\end{figure*}
The left plot in Figure~\ref{fig:results} shows the performance of FSE using the parameters from Table~\ref{tab:fse_parameters} compared to OMP-G and Janssen's method on the training data set GRID. It can be seen that FSE always gives better results than Janssen's method and also better results than OMP-G except for a clipping level of $0.9$. However, for this slight clipping, the reconstruction quality is already $35.57$ dB and the difference to OMP-G with $35.91$ dB is very small and therefore audibly not noticeable. On the average, FSE outperforms Janssen's method by $2.54$ dB and OMP-G by $1.31$ dB on the GRID test data set.

For the comparative evaluation, FSE, OMP-G, and Janssen's method are tested on the SISEC data set which is independent of the GRID data set. FSE uses the parameters from Table~\ref{tab:fse_parameters}. As it can be seen in Figure~\ref{fig:results}, OMP-G gives better results than Janssens's method for clipping levels of $0.7$ and higher. For clipping levels below $0.7$, Janssen's method performs better than OMP-G. FSE shows a similar behavior like OMP-G for lower clipping levels, however, for clipping levels higher than $0.6$, FSE gives a significantly better reconstruction quality than OMP-G and Janssen's method. On the average, FSE outperforms OMP-G by $1.01$ dB and Janssen's method by $1.76$ dB on the SISEC test data set.
FSE not only outperforms OMP-G in terms of SNR, it also requires less computation time. As shown in Table~\ref{tab:time_measurements}, declipping using FSE is on average around $100$ times faster than by OMP-G. Janssen's method may outperform FSE and OMP-G in computation time, however, it also gives the lowest reconstruction quality of the considered declipping algorithms.

\begin{figure}
	\centering
	\def\svgwidth{\columnwidth}
	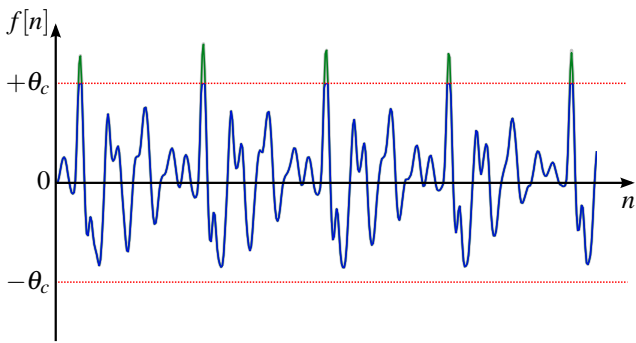
	\vspace*{-0.5cm}
	\caption{Reconstruction of a distorted audio signal: Clipped signal (blue), original signal (light gray), and reconstructed signal (green).}
	\label{fig:rec_signal}
	\vspace*{-0.3cm}
\end{figure}

The reconstruction of a clipped audio signal is illustrated in Figure~\ref{fig:rec_signal}. Here, FSE is applied to a clipped speech signal where the blue curve represents the clipped signal, light gray is the original signal, and green is the reconstructed signal. One can see that the reconstructed samples are very close to the original samples.

\vspace*{0.2cm}
\section{Conclusion}
\label{sec:conclusion}
\vspace*{0.1cm}

In this paper, a reconstruction algorithm which is normally used in image processing is presented and adapted to be able to restore audio signals which are distorted from clipping. It has been shown that the proposed algorithm gives a good reconstruction quality. Comparative evaluation experiments have shown that compared to other state-of-the-art declipping algorithms, gains in SNR of up to $3.52$ dB and an average gain of $1.01$ dB are possible.

\newpage
\small
\bibliographystyle{ieeetr}
\bibliography{literature}

\end{document}